\documentclass[journal=acscii]{achemso}
\usepackage[T1]{fontenc} 
\usepackage[version=3]{mhchem}
\usepackage{amsmath}
\usepackage{amssymb}
\usepackage{graphicx}
\usepackage{subcaption}
\usepackage{color}
\usepackage{dcolumn}
\usepackage{bm}
\usepackage{miller}
\usepackage[utf8]{inputenc}
\usepackage{nicefrac}
\usepackage{multirow}
\usepackage{makecell}
\usepackage{nameref}
\usepackage{hyperref}
\usepackage{pdfpages}
\hypersetup{
    colorlinks=true,
    linkcolor=blue,
    filecolor=magenta, 
    urlcolor=cyan,
    citecolor=blue
}

\definecolor{orange}{RGB}{255, 108, 12}

\graphicspath{ {Figures/} }
\author{Victor Venturi}
\affiliation{Department of Mechanical Engineering, Carnegie Mellon University, Pittsburgh, Pennsylvania 15213, USA}
\author{Holden Parks}
\affiliation{Department of Mechanical Engineering, Carnegie Mellon University, Pittsburgh, Pennsylvania 15213, USA}
\author{Zeeshan Ahmad}
\affiliation{Department of Mechanical Engineering, Carnegie Mellon University, Pittsburgh, Pennsylvania 15213, USA}
\author{Venkatasubramanian Viswanathan}
\affiliation{Department of Mechanical Engineering, Carnegie Mellon University, Pittsburgh, Pennsylvania 15213, USA}
\alsoaffiliation[phys]{Department of Physics, Carnegie Mellon University, Pittsburgh, Pennsylvania 15213, USA}
\email{venkvis@cmu.edu}

\title[An \textsf{achemso} demo]
  {Machine Learning Enabled Discovery of Application Dependent Design Principles for Two-dimensional Materials}

\begin{document}

\begin{tocentry}
\includegraphics[height=3.7cm]{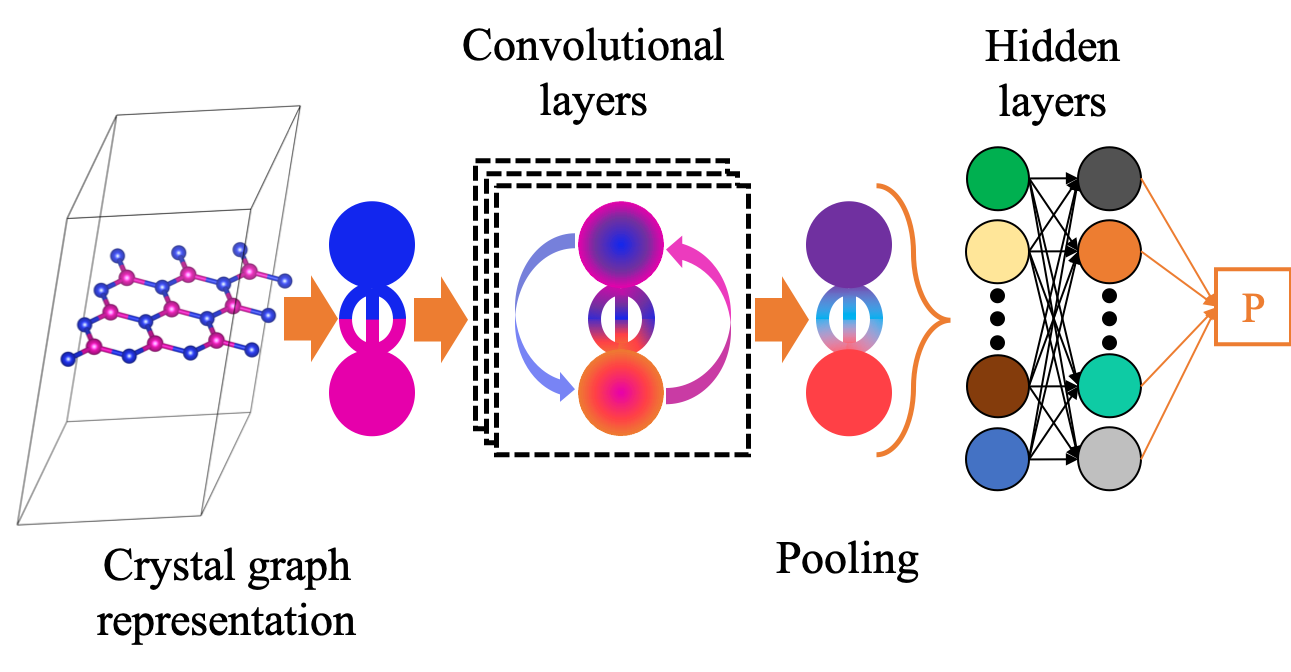}
\\
Synopsis: We use crystal graph convolutional neural networks (CGCNN) to screen nearly 45,000 monolayer 2D materials for applications in: composites and photovoltaics. We identify underlying compositional and structural design rules that govern the performance of these structures in these applications.
\end{tocentry}

\begin{abstract}

The unique electronic, mechanical, and magnetic properties of two-dimensional (2D) materials make them promising next-generation candidates for a variety of energy applications. MXenes, a relatively new class of 2D materials, have been investigated as components in mechanically robust composites, while 2D perovskites have shown exceptional promise for solar cell and water splitting. The large-scale search for high-performing candidate 2D materials is limited to calculating a few simple descriptors, usually with first-principles density functional theory calculations.  In this work, we alleviate this issue by extending and generalizing crystal graph convolutional neural networks to systems with planar periodicity, and train an ensemble of models to predict thermodynamic, mechanical, and electronic properties. To demonstrate the utility of this approach, we carry out a screening of nearly 45,000 structures for two largely disjoint applications: namely, mechanically robust composites and photovoltaics. An analysis of the uncertainty associated with our methods indicates the ensemble of neural networks is well-calibrated and has errors comparable with those from accurate first-principles density functional theory calculations. The ensemble of models allows us to gauge the confidence of our predictions, and to find the candidates most likely to exhibit effective performance in their applications. Since the datasets used in our screening were combinatorically generated, we are also able to investigate, using an innovative method, structural and compositional design principles that impact the properties of the structures surveyed and which can act as a generative model basis for future material discovery through reverse engineering. Our approach allowed us to recover some well-accepted design principles: for instance, we find that hybrid organic-inorganic perovskites with lead and tin tend to be good candidates for solar cell applications. Similarly, we find that titanium based MXenes usually have high stiffness coefficients, but, interestingly, the other members of the group 4 of the periodic table, namely, zirconium and hafnium, also contribute to increasing the mechanical strength of these structures. In the case of all-inorganic perovskites, we discover that those with favorable band gaps have scandium, zirconium, or hafnium occupied A-sites, with chromium, scandium, vanadium, or niobium on the B-sites; combinations that have not been deeply studied in the field of photovoltaics and thus open up paths for further investigation.  We open-source the code-base and datasets to spur further development in this space.

\end{abstract}


\section{Introduction}\label{intro}

\begin{figure}[htbp]
    \centering
    \includegraphics[scale=.6]{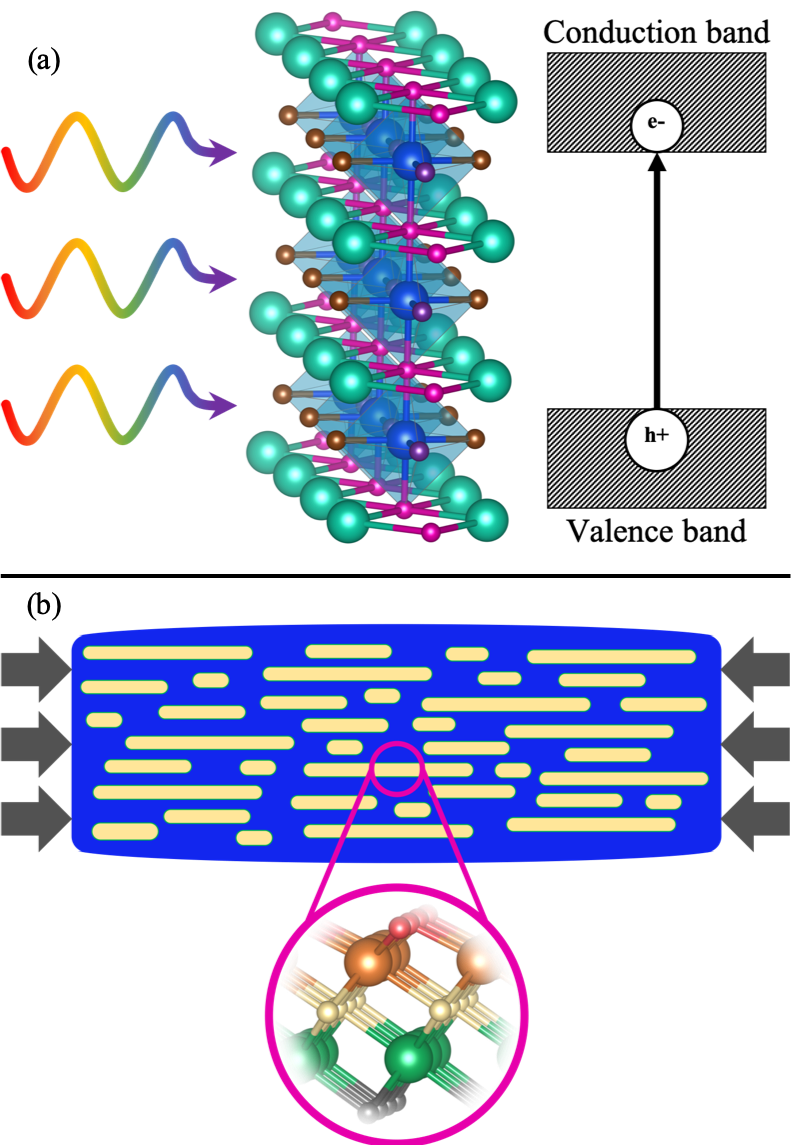}
    \caption{ Illustration of 2D materials applications: (a) 2D perovskites whose bandgaps can be the range of 1.5-3 eV have been demonstrating promise in the field of photovoltaics, ; (b) MXenes can be used in composites to increase their mechanical strength, here indicated by the material's response to an external mechanical stress (in grey).}
    \label{fig:apps_combined}
\end{figure}

Two-dimensional (2D) materials have emerged as attractive candidates for energy applications due to their unique electronic, mechanical, chemical, optoelectronic and magnetic properties~\cite{Novoselov2016, Zhang20162D, Quesnel2015, Ge2016}.  Among their different prototype structures, MXenes have been explored for applications in battery electrodes, water purification, catalysis, lubrication etc.~\cite{Pang2019,Er2014,Chaudhari2017}. Their structure and composition allows the careful tuning of properties for these applications~\cite{Anasori2017}. Most device applications of 2D materials require mechanical integration with the substrate, promoting an interest in their mechanical properties. MXenes are known to be mechanically stronger compared to other 2D materials resulting in applications in protective coatings, composites and membranes~\cite{Lipatov2018}. 2D materials offer a new way of tuning the properties of their 3D counterparts like band gaps through exfoliation~\cite{Mounet2018}. 2D counterparts of perovskites have shown promise for solar cell applications~\cite{Xiao2016} (Figure \ref{fig:apps_combined}).

The existence of a variety of 2D structures and atoms to populate their sites imply that a purely experimental or computational approach based on first-principles calculations to identify materials for desired applications is infeasible. For example, an MXene structure of the form \ce{M_{n+1} X_n T_x} (X=C, N) when provided with $m$ metal possibilities, $t$ functional group possibilities combinatorially explodes: the upper bound on the number of materials would be $\sim 2^n m^{n+1} t^{x}$; for example, the case, $n=2$ (i.e. M$_3$X$_2$) with $m=10$ metal possibilities and $t=3$ different terminations (e.g. F, OH, O) on either side of the structure would yield approximately $35,000$ materials, while just doubling the number of metal options from 10 to $m=20$ gives $\sim 280,000$ possibilities. With the generation of massive amounts of materials data~\cite{Himanen2019data}, data-driven techniques offer a new avenue to tackle this problem~\cite{Agrawal2016materials}. Data-driven methods have shown the promise of not only furthering our fundamental understanding of materials~\cite{Umehara2019analyzing} but also provide a platform for performing large scale computational screening through the development of accurate structure-property relationships~\cite{coley2017, evans2017predict, sendek2017holistic, ahmad2018}. Machine learning methods can bypass the use of expensive first-principles calculations and help accelerate the often time consuming discovery and optimization of materials for various applications~\cite{pilania2013accelerating, fujimura2013method, Kim2018machine}.

Recently, graph convolution based machine learning models have shown promising generalization capability for predicting the properties of crystals and molecules~\cite{xie2017crystal,ahmad2018,Schtt2018Schnet,Chen2019graph}. These methods encode the structure of a material as a graph based on the position and coordination of atoms, thus circumventing the use of carefully handcrafted or engineered structural features. This enables them to be used in a variety of applications. Here, we extend crystal graph convolutional neural networks (CGCNN) to study materials with planar periodicity. Using 100 different, randomly generated training sets, we trained an ensemble of CGCNN models to predict thermodynamic, mechanical, and electronic properties. The ensemble of neural networks shows errors comparable to those from highly accurate first-principles calculations, such as density functional theory (DFT), as discussed in the Supporting Information. We use this ensemble to screen $\sim 45,000$ 2D monolayer materials with focus on mechanically strong MXenes ($c_{11}, c_{22} \geq 175$ N/m) and in perovskites whose band gaps fall within an acceptable range ($[1.5, 3]$ eV) for solar cell applications. The two applications chosen are quite different from each other and aims to demonstrate the generalizability of our model predictions.  With these models, We recover some well-accepted design rules: for instance, hybrid organic-inorganic perovskites with either tin or lead as metal components are useful for photovoltaics, as well as fomamidinium, imidazolium, or azetidinium fulfilling the role of organic cations. Similarly, we find that titanium based MXenes tend to be mechanically robust, however, incorporating the other main elements of group 4 of the periodic table, zirconium and hafnium, also aids in increasing the stiffness of this class of structures.  Interestingly, we also identify some lesser investigated principles: all-inorganic perovskites whose band gaps are most likely to fall within a desirable range for solar cell applications have zirconium, niobium, hafnium, scandium, or vanadium as metal components.

\section{Methods}

\subsection{Databases}\label{datasets}

The present work utilizes computational data and material structures from (1) the Computational 2D Materials Database (C2DB) \cite{Haastrup_2018}, (2) a database of hybrid organic-inorganic perovskites generated by \citealt{Kim2017} (HOIP), (3) a database of cubic perovskites generated by \citealt{Castelli2012} (Castelli), and (4) a database of 2D MXenes generated by~\citealt{anant} (aNANt). 

We primarily used data from the C2DB database to train the CGCNN model. As of August 2019, C2DB consists of over 3500 structural, thermodynamic, elastic, electronic, magnetic, and optical properties calculated using density functional theory (DFT). Each structure was combinatorially generated from a series of prototype structures that differ in space group, stoichiometry, and thickness. Some example prototypes include BN (space group P$\bar{3}$m2), BiI$_3$ (P$\bar{3}$m1), or PbSe (P4/mmm). DFT calculations were performed using the Perdew, Burke, and Ernzerhof (PBE) exchange correlation functional~\cite{GGAPBE} in the projector augmented wave (PAW) code GPAW~\cite{enkovaara2010electronic}. The stability of each material is evaluated by predicting enthalpy of formation, the elastic constants, and the phonon frequencies. If a material is stable, its electronic structure and other properties, such as polarizability, are also calculated. The most relevant properties for screening 2D MXenes and perovskites are the heat of formation; bandgap; and the $c_{11}$, $c_{12}$, and $c_{22}$ components of the elastic tensor. 

Using the CGCNN models trained from the C2DB data, we predict the heat of formation, bandgap, and in-plane elastic tensor components of approximately $20,000$ 2D perovskites and $25,000$ 2D MXenes. The perovskite structures were taken from two sources: the HOIP dataset\cite{Kim2017}, and the Castelli database\cite{Castelli2012}. The HOIP database contains 1,346 structures that were combinatorially-generated from a series of 135 prototypes. The perovskite prototypes were obtained using the minima-hopping method outlined by Goedecker~\cite{Goedecker2004}. Each structure has stoichiometry ABX$_3$, where A is one of 16 organic cations, B is one of $\{$Ge, Pb, Sn$\}$, and $X$ is one of $\{$F, Cl, Br, I$\}$. The Castelli database contains nearly 19,000 cubic perovskites. The structures were generated combinatorially with each perovskite having stoichiometry ABX$_3$, where A and B are each one of 52 different metals and X$_3$ is one of 7 different anion groups. Both databases contain only bulk structures. To create 2D lattices we exfoliate the bulk structures to generate a (001) monolayer. 

The 2D MXenes structures are taken from the aNANt database\citealt{anant}. This database contains combinatorially-generated 23,870 MXenes with five-layer structures of the form T-M-X-M'-T' (that is, the T/T' occupy the outermost layers in the structure), where T and T' are each one of 14 termination functional groups, M and M' are each one of 11 early transition metals, and X is one of $\{$C, N$\}$.


\subsection{Model Training}

\begin{figure}[htbp]
    \centering
    \includegraphics[width=.75\textwidth]{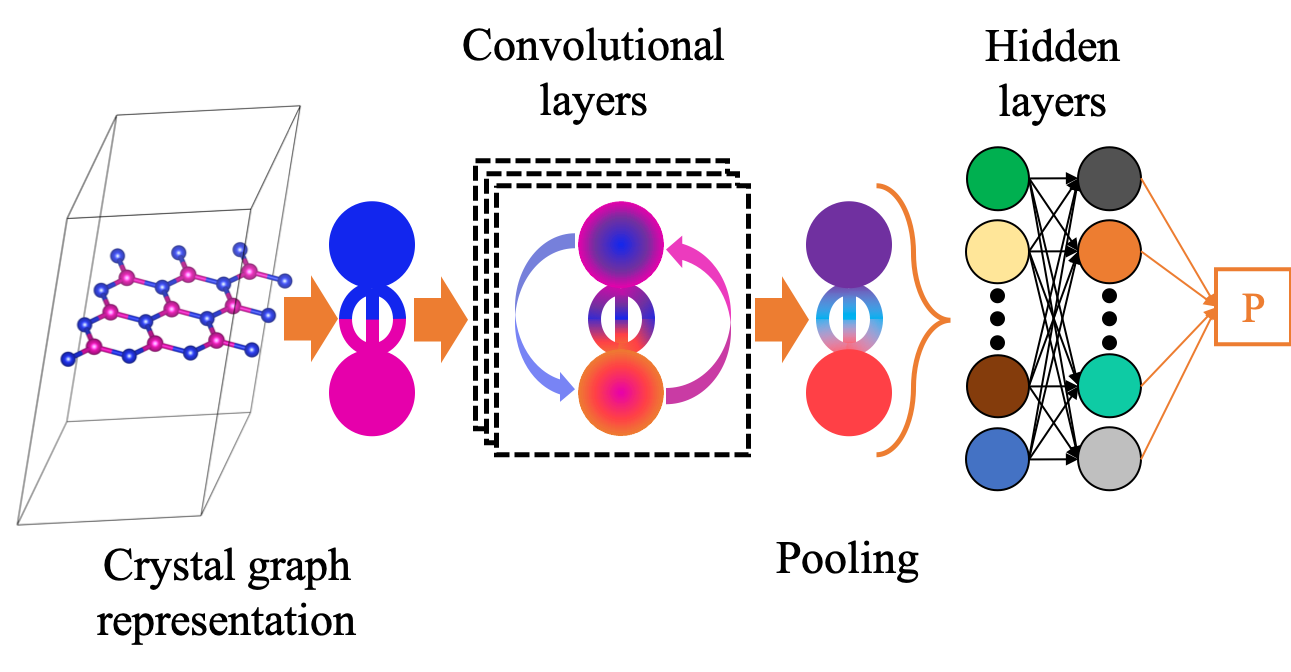}
    \caption{A simple representation of the CGCNN architecture. The atomic structure is first converted into its crystal graph representation. This graph is then passed as input for the convolutional layers, where the atomic feature vectors are updated based on neighbor information and bond lengths. Next, a pooling function is employed to produce an overall vector representation of the crystal, guaranteeing invariance with respect to number of primitive unit cells used in the creation of the original structure. Finally, a set of fully connected hidden layers maps the simplified vector-represented crystal structure to the property of interest.
    }
    \label{fig:cgcnn_schematics}
\end{figure}

In order to screen the $\sim 20,000$ perovskites and $\sim 24,000$ MXenes 2D monolayer structures, as well as to uncover the underlying design principles for their respective applications, a technique that can predict properties accurately at a computational cost much lower than DFT is required. We use the Crystal Graph Convolutional Neural Network (CGCNN) framework~\cite{xie2017crystal} as a surrogate technique for predicting material properties. This method provides the accuracy of DFT calculations (discussed in the Supporting Information) but at a fraction of the associated computational cost: while it can take up to 500 CPU hours to compute the $c_{11}$ coefficient for one structure with DFT, CGCNNs can predict the same property for roughly 25,000 structures in under 20 GPU minutes once trained. This framework has been successfully used in a variety of applications, from selecting solid electrolyte candidates \cite{ahmad2018} to screening catalytic materials \cite{zulissi2019}. At the foundation of the CGCNN is the undirected multigraph representation of the crystal structure, in which nodes represent atoms by their respective features, and edges encode interatomic bond distances~\cite{xie2017crystal}. Iterative convolution layers update atomic feature vectors based on neighbor information. A simplified depiction of the CGCNN can be seen in Figure \ref{fig:cgcnn_schematics}. 

After optimizing the network architecture (see Supporting Information), we used an ensemble of 100 CGCNN models, each trained on a random set of 70\% of the C2DB data to predict the properties of interest: band gap, $\log(c_{11})$, $\log(c_{22})$, $c_{12}$, conduction band minimum (CBM), valence band maximum (VBM), and heat of formation ($H_{\text{form}}$).

\section{Results and Discussion}

\subsection{Structure Screening}\label{screening}

\begin{figure}[H]
    \centering
    \includegraphics[width=1\textwidth]{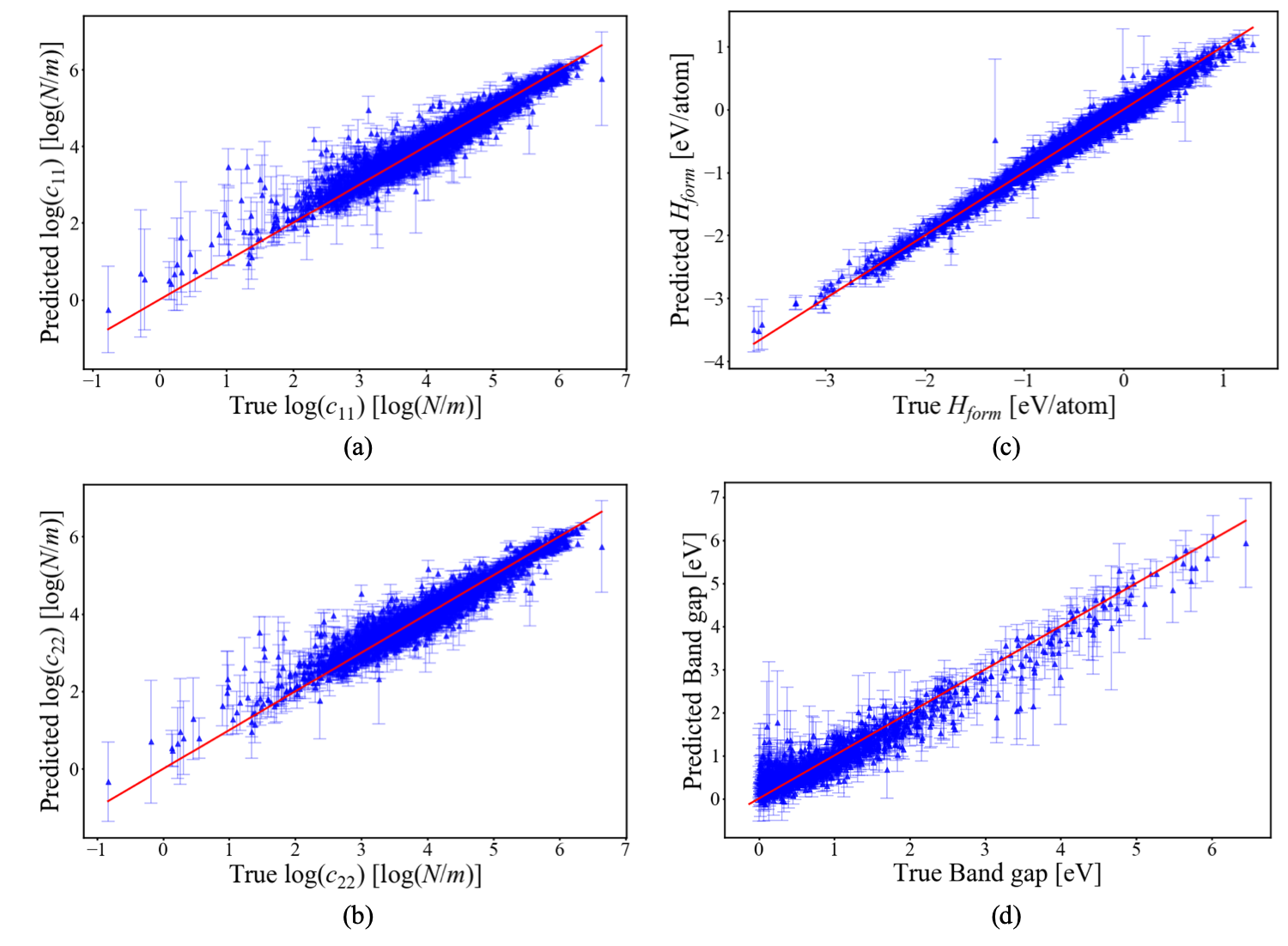}
    \caption{Comparison of predicted and DFT-calculated (a)  $\log(c_{11})$, (b) $\log(c_{11})$, (c) $H_{\text{form}}$, and (d) band gap on C2DB data. The predictions are made using an ensemble of 100 CGCNN models, each trained on randomly selected training data.}
    \label{fig:c2db}
\end{figure}

In order to evaluate the accuracy of the ensemble of 100 models, we used them to predict the properties of the all the structures in the C2DB database. The results for the ensemble predictions of some of the main properties of interest can be seen in the parity plots shown in Figure \ref{fig:c2db}. A discussion on the usefulness of the uncertainty quantification of models can be found in the Supporting Information.

As discussed in the \nameref{intro}, we screened MXenes in search of structures that are strong mechanically, with both $c_{11}, c_{22} \geq 175$ N/m, thus exceeding those of graphene oxide \cite{Lipatov2018}, and perovskites whose bandgaps fall in the range $[1.5, 3]$ eV, appropriate for solar cell applications. Since it is also required that these structures be stable, as well as synthesizable, our filtering procedure included the additional requirement that $H_{\text{form}} \leq -2$ eV/atom for MXenes and inorganic perovskites, and $H_{\text{form}} \leq -0.5$ eV/atom for hybrid organic-inorganic perovskites. The difference in treatment for the latter stems from the fact that hybrid perovskites are known to be relatively less stable than their inorganic counterparts \cite{meng2018}. These threshold values of $H_{\text{form}}$ were chosen as they represent the average of the lowest heats of formation of the structures in their respective datasets.

We quantify the confidence of our predictions for a given structure $s$ by its $c$-value (confidence value)~\cite{greg2017} representing the fraction of models in the ensemble that predict structure $s$ to be useful for the application, based on the aforementioned criterion. It is calculated in the following manner:
\begin{equation}\label{c_val_intro}
    c(s) = \frac{1}{N} \sum\limits_{i=1}^N \mathcal{M}_i(s),
\end{equation}
where $N=100$ is the number of models used and 
\begin{equation}
    \mathcal{M}_i(s) =
    \begin{cases}
        & 1 \text{ if the $i$th model predicts the given structure $s$ to be useful} \\
        & 0 \text{ otherwise}
    \end{cases}
\end{equation}
This enables us to determine the 2D structures with the highest likelihood of being useful for their applications, as shown in Table \ref{tab:top_structs}. It is important to note, however, that the training sets for the band gap prediction models contained only materials with non-zero band gap, meaning that a further metallic versus insulator filtering, with subsequent update of the $c$-values, is needed. The reason for this is that, when given a conducting material, they predict a positive band gap, since they have only been trained on insulators or semiconductors.
\begin{table}[htp]
    \centering
    \begin{tabular}{c||cccccc}
        \hline \hline
        & Structure & c-value & \makecell{$\langle H_{\text{form}} \rangle$  \\\text{[eV/atom]}} & \makecell{$\langle c_{11} \rangle$\\\text{[N/m]}} & \makecell{$\langle c_{22} \rangle$\\\text{[N/m]}}& \makecell{$\langle \text{band gap}\rangle$ \\\text{[eV]}}\\
        \hline\hline
        \multirow{5}{*}{\rotatebox[origin=c]{90}{MXenes}}
        & Hf*O-N-Hf*O & 1.0 & -2.62 & 273.88 & 261.60 & -\\
        & Sc*O-N-Hf*O & 1.0 & -2.64 & 210.15 & 224.16 & -\\
        & Sc*F-N-Hf*O & 1.0 & -2.63 & 240.97 & 220.28 & -\\
        & Hf*O-C-Zr*F & 1.0 & -2.23 & 266.41 & 249.99 & -\\
        & Hf*O-N-Zr*Cl & 1.0 & -2.16 & 211.11 & 227.78 & -\\
        \hline\hline
        \multirow{5}{*}{\rotatebox[origin=c]{90}{\makecell{Inorganic \\Perovskites}}}
        & \ce{NbZrO3}; A=Zr, B=Nb & 0.98 & -2.48 & - & - & 2.28\\
        & \ce{HfVO3}; A=Hf, B=V & 0.98 & -2.33 & - & - & 2.33\\
        & \ce{MoZrO3}; A=Zr, B=Mo & 0.94 & -2.25 & - & - & 2.38\\
        & \ce{HfNbO3}; A=Nb, B=Hf & 0.94 & -2.20 & - & - & 1.99\\
        & \ce{NbTiO3}; A=Nb, B=Ti & 0.92 & -2.18 & - & - & 2.08\\
        \hline\hline
        \multirow{5}{*}{\rotatebox[origin=c]{90}{\makecell{Organic\\Perovskites}}}
        & \ce{C3H5F7N2Sn2}; A=\ce{C3H5N2} & 0.99 & -1.21 & - & - & 2.38\\
        & \ce{C3H8F7NPb2}; A=\ce{C3H8N} & 0.94 & -1.13 & - & - & 2.67\\
        & \ce{C3H5F7N2Pb2}; A=\ce{C3H5N2} & 0.93 & -1.19 & - & - & 2.63\\
        & \ce{C2H7F7N2Sn2}; A=\ce{CH3C(NH2)2} & 0.92 & -1.37 & - & - & 2.67\\
        & \ce{CH5F7N2Sn2}; A=\ce{HC(NH2)2} & 0.91 & -1.60 & - & - & 2.36\\
        \hline\hline
    \end{tabular}
    \caption{\label{tab:top_structs} Materials with highest five $c$-values. For MXenes, `*' indicates a bond between a metallic atom and a termination. For the perovskites, the site occupations have been specified for clarity. Values reported are the mean of the ensemble predictions.}
\end{table}

\subsection{Identifying Design Principles}\label{identifyDP}

To uncover the compositional and structural commonalities of useful candidates, we applied an analogous concept to study the design principles that can increase the $c$-values of different MXene and perovskite materials. 
First, for each dataset used, we establish the following functions of the design principle (DP): the subset of all structures satisfying the DP, $\mathcal{D}_{\text{DP}} = \{\text{structures that satisfy the DP}\}$; the proportion of the dataset that contains the DP, $P_{\text{DP}} = N_{\text{DP}}/N_{\text{dataset}}$, where $N_{\text{DP}} = |\mathcal{D}_{\text{DP}}|$ is the cardinality of set $\mathcal{D}_{\text{DP}}$ (the number of elements in this set), and $N_{\text{dataset}}$ is the total number of structures in the dataset; and the average of $c$-values of all structures in $\mathcal{D}_{\text{DP}}$
\begin{equation*}
    c_{\text{DP}} = \frac{1}{N_{\text{DP}}}\sum\limits_{s \in \mathcal{D}_{\text{DP}}} c(s),
\end{equation*}
which can be interpreted as the chance of an arbitrary model predicting that a random structure in $\mathcal{D}_{\text{DP}}$ is a useful candidate.

\begin{figure}[htbp]
    \centering
    \includegraphics[width=1\textwidth]{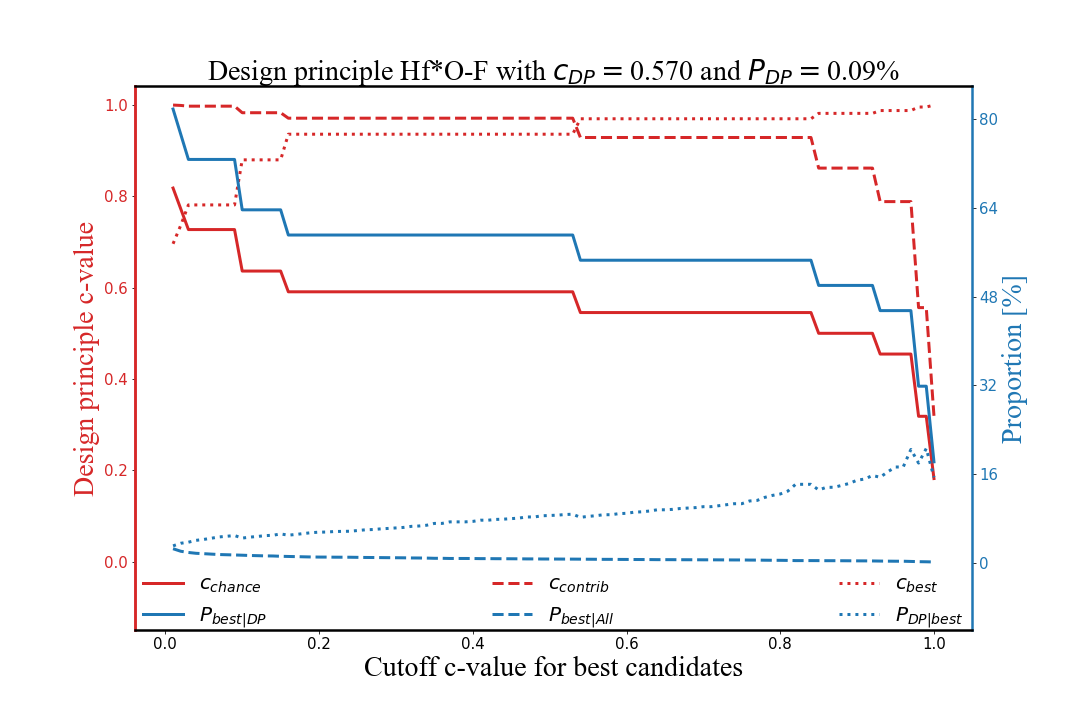}
    \caption{One of the top design principles for MXenes: hafnium (Hf) bonded with oxygen (O) termination, and with a fluorine (F) termination bonded to any other metal. The likelihood that any arbitrary model predicts that a random material satisfying this DP is a useful candidate is of $\sim 57$\%, as indicated by $c_{\text{DP}}$. Choosing $c_{\text{cut}}=1.0$ shows that the chance of a structure satisfying the DP being among the best candidates is of nearly $c_{chanceDP} = P_{\text{best|DP}}\sim20\%$, and these candiates contribute to approximately 30\% of $c_{\text{DP}}$, as shown by $c_{\text{contrib}DP}$. Note that, for this specific DP, $P_{\text{DP|best}}$ almost steadily increase with the value of $c_{\text{cut}}$, indicating that, the more confident we want to be in our set of useful candidates, in general, the more prevalent this DP becomes in this $\mathcal{B}$ set.}
    \label{fig:mxene_top_DP}
\end{figure}

\begin{figure}[htbp]
    \centering
    \includegraphics[width=1\textwidth]{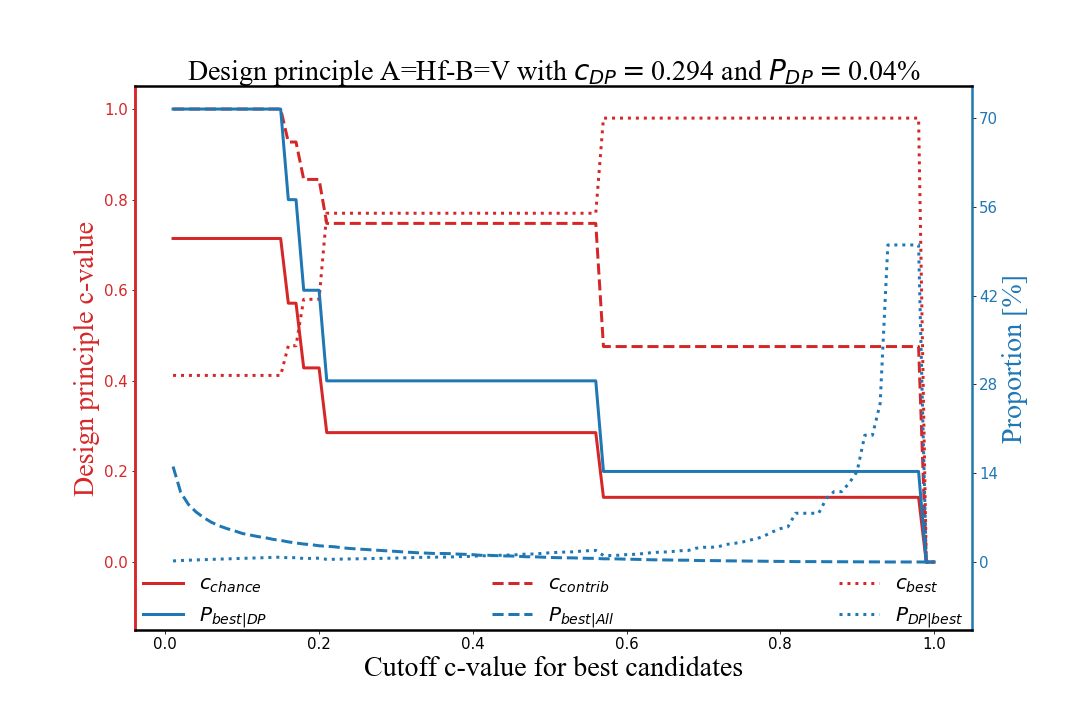}
    \caption{One of the top design principles for inorganic perovskites: A-site occupied by hafnium (Hf), with vanadium (V) in the B-site. The likelihood that any arbitrary model predicts a random material satisfying this DP is a useful candidate is of $\sim29\%$, as indicated by $c_{\text{DP}}$. Choosing $c_{\text{cut}}=0.80$ shows that the chance of a structure from set $\mathcal{D}_{\text{DP}}$ to be among the best candidates is of $c_{\text{chance}DP} = P_{\text{best|DP}}=14\%$, and these candidates contribute to approximately 50\% of $c_{\text{DP}}$, as shown by $c_{\text{contrib}DP}$. Additionally, for this specific DP, $P_{\text{DP|best}}$ almost steadily increases with the value of $c_{\text{cut}}$, indicating that, the more confident we want to be in our set of useful candidates, in general, the more prevalent this DP becomes in this $\mathcal{B}$ set. Finally, an examination of the behavior of $c_{\text{best}}$ reveals that, for values of the cutoff $c_{\text{cut}}>0.6$, all of the structures in set $\mathcal{B}\cap\mathcal{D}_{\text{DP}}$ have a $c$-value of nearly 1.0.}
    \label{fig:castelli_top_DP}
\end{figure}

\begin{figure}[htbp]
    \centering
    \includegraphics[width=1\textwidth]{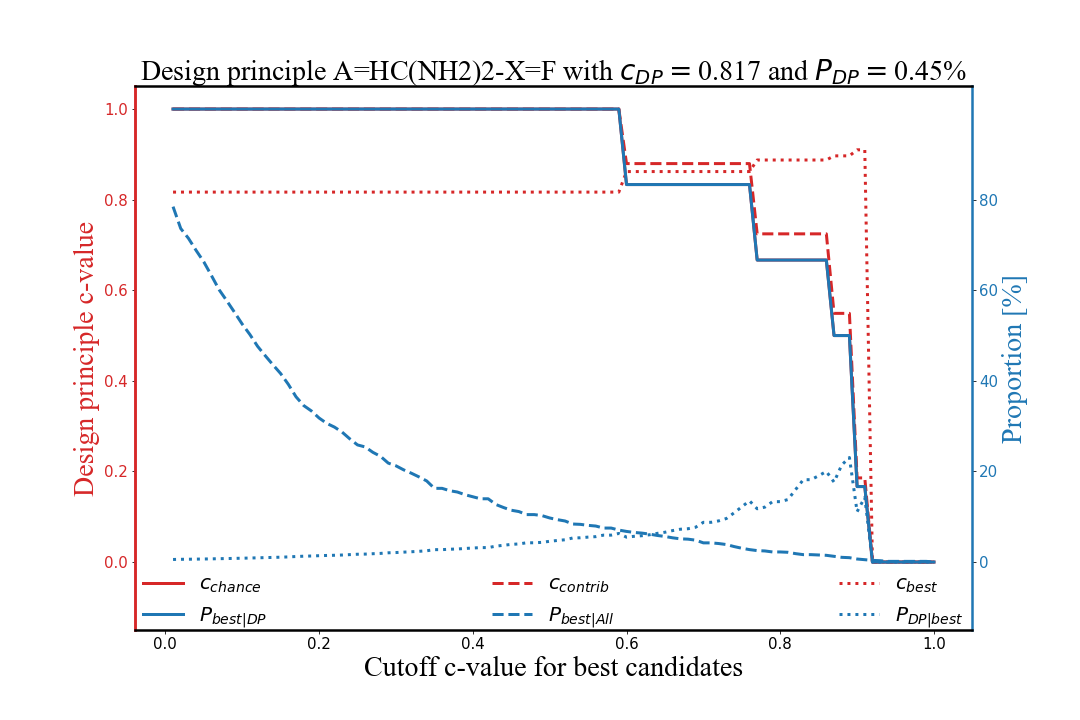}
    \caption{Top design principle for Khazana perovskites: A-site occupied by fomamidinium (\ce{HC(NH2)2}), with fluorine (F) in all X-sites. The likelihood that any arbitrary model predicts a random material satisfying this DP is a useful candidate is of $\sim82\%$, as indicated by $c_{\text{DP}}$. Choosing $c_{\text{cut}}=0.80$ shows that the chance of a structure from set $\mathcal{D}_{\text{DP}}$ to be among the best candidates is of $c_{\text{chance}DP} = P_{\text{best|DP}}=\sim65\%$, and these candidates contribute to approximately $\sim70\%$ of $c_{\text{DP}}$, as shown by $c_{\text{contrib}DP}$.}
    \label{fig:khazana_top_DP}
\end{figure}

Next, we introduce a minimum threshold, $c_{\text{cut}}$, to distinguish the best candidate structures from the others. The subset composed of these materials can be expressed as a function of $c_{\text{cut}}$ as 
$\mathcal{B}(c_{\text{cut}}) = \{ \text{structures with $c$-value} \geq c_{\text{cut}}\}$.
From this definition, we can examine how the presence of a specific design rule in a material influences its existence among the best candidates in set $\mathcal{B}(c_{\text{cut}})$, as well as the $c$-values of these structures.
For this purpose, we will define a few quantities, all functions of $c_{\text{cut}}$. One of the simplest indicators that a given DP is effective at making a structure useful for the application in a combinatorically generated dataset is the proportion of the set of best candidates $\mathcal{B}$ that is comprised of materials satisfying the DP, $P_{\text{DP|best}} = |\mathcal{B}\cap\mathcal{D}_{\text{DP}}|/|\mathcal{B}|$, and how it compares with $P_{\text{DP}}$. Additionally, it is helpful to examine the difference between the likelihood of a random material being amongst the best candidates $P_{\text{best|All}} = |\mathcal{B}|/N_{\text{dataset}}$ and the chance of that happening given that the structure contains the DP, $c_{\text{chance}DP} = P_{\text{best|DP}} = |\mathcal{B}\cap\mathcal{D}_{\text{DP}}|/|\mathcal{D}_{\text{DP}}|$. Besides these quantities, it is also important to measure our confidence in these candidates, members of $\mathcal{B}\cap\mathcal{D}_{\text{DP}}$, by averaging their $c$-values:
\begin{equation*}
    c_{\text{bestDP}} = \frac{1}{|\mathcal{B}\cap\mathcal{D}_{\text{DP}}|}\sum\limits_{s\in \mathcal{B}\cap\mathcal{D}_{\text{DP}}} c(s).
\end{equation*}
Note that, by construction, $c_{\text{bestDP}}$ is a monotonically increasing function of $c_{\text{cut}}$ while the set $\mathcal{B}\cap\mathcal{D}_{\text{DP}}$ is not empty. Finally, although redundant with all previously described measures, we also studied, for completeness, how the elements from $\mathcal{B}\cap\mathcal{D}_{\text{DP}}$ contribute to $c_{\text{DP}}$:
\begin{equation*}
    c_{\text{contribDP}} = \frac{\sum\limits_{s\in \mathcal{B}\cap\mathcal{D}_{\text{DP}}}c(s)}{\sum\limits_{s'\in\mathcal{D}_{\text{DP}}}c(s')} = \frac{|\mathcal{B}\cap\mathcal{D}_{\text{DP}}|c_{\text{bestDP}}}{|\mathcal{D}_{\text{DP}}|c_{\text{DP}}} = P_{\text{best|DP}}\frac{c_{\text{bestDP}}}{c_{\text{DP}}}.
\end{equation*}
Since the higher the value of the cutoff, the fewer elements are in the set $\mathcal{B}\cap\mathcal{D}_{\text{DP}}$, both $c_{\text{contribDP}}$ and $c_{\text{chanceDP}}$ are monotonically decreasing with $c_{\text{cut}}$. A full dependency of all these variables with the value of the cutoff $c_{\text{cut}}$, for chosen design principles, can be seen in Figures \ref{fig:mxene_top_DP}, \ref{fig:castelli_top_DP}, and \ref{fig:khazana_top_DP}. Note that, for all three of the design rules chosen, $P_{\text{DP|best}}$ is almost a monotonically increasing function of $c_{\text{cut}}$, indicating that, the more confident one wants to be on the $\mathcal{B}$ set, the more predominant these design principles become in this set. Additionally, prior to the value of $c_{\text{cut}}$ for which none of the materials in $\mathcal{B}$ contains the design principles, the chance of finding a member of $\mathcal{B}$ among the set $\mathcal{D}_{\text{DP}}$ is of roughly 15\% for all three design principles. (Note: values for $c_{\text{cut}}=0$ omitted in the interest of ease of graphical visualization.)

This approach of understanding the effect of design principles is best suited for combinatorially generated datasets. Therefore, we used it to study all of the data mentioned in the \nameref{datasets} Section, including that from HOIP \cite{Kim2017}. We constructed a list of all possible design principles using the same combinatorics applied in the creation of the respective datasets. These design principles were then ordered by highest to lowest $P_{\text{DP|best}} / P_{\text{DP}}$  ratio at a cutoff value of $c_{\text{cut}}=0.95$ for MXenes, and $c_{\text{cut}}=0.80$  for both inorganic and organic perovskites. Our choice for cutoff values was guided by the results from Table \ref{tab:top_structs}: we wanted to make sure that the set of best candidates $\mathcal{B}$ was sizeable enough for a meaningful analysis of the design principles. Furthermore, for the MXenes, we excluded the design rules whose $P_{\text{DP}} \leq 0.008\%$ due to their high specificity. We would like to note the following interesting equality, which establishes a relationship between the two most intuitive criteria of gauging the effectiveness of a given DP discussed previously:
\begin{equation}
    \frac{P_{\text{DP|best}}}{P_{\text{DP}}} = \frac{|\mathcal{B}\cap\mathcal{D}_{\text{DP}}|}{|\mathcal{B}|}\frac{N_{\text{dataset}}}{|\mathcal{D}_{\text{DP}}|} = \frac{|\mathcal{B}\cap\mathcal{D}_{\text{DP}}|}{|\mathcal{D}_{\text{DP}}|}\frac{N_{\text{dataset}}}{|\mathcal{B}|} = \frac{P_{\text{best|DP}}}{P_{\text{best|All}}}
\end{equation}
The results of this analysis are shown in Table \ref{tab:DP_table_res}. We have also chosen one of the top design principles for MXenes (Figure \ref{fig:mxene_top_DP}), inorganic (Figure \ref{fig:castelli_top_DP}), and organic perovskites (Figure \ref{fig:khazana_top_DP}) to represent the dependency between the metrics discussed above and the cutoff $c_{\text{cut}}$, which determines the minimum confidence level of the structures in the set of best candidates $\mathcal{B}$. 

\subsection{Interpreting Design Principles}\label{interpretDP}

Our study was able to identify some known design rules: for example, we see that titanium (Ti) based MXenes tend to have high stiffness coefficients, as suggested by Figure 4 and Table 1 in~\citealt{Anasori2017}. Interestingly, however, we discovered that the other main elements of group 4 of the periodic table, namely, zirconium (Zr) and hafnium (Hf), can also increase the mechanical strength of this class of materials, as long as these elements are bonded with oxygen, and the opposite side of the monolayer is either oxygen or fluorine terminated.

Similarly, our model was able to recognize that, in order for hybrid organic-inorganic perovskites to have a band gap in the range of $[1.5, 3]$ eV, the  B-sites should be occupied by either lead (Pb) or tin (Sn), a relatively well-known design principle in the photovoltaics community. \cite{Hao2014leadfree, Noel2014, Yi2019, toshniwal2017} We have also found that the organic A-sites should be composed of fomamidinium \ce{(HC(NH2)2)}, imidazolium (\ce{C3H5N2}), or azetidinium(\ce{C3H8N}). Curiously, for these hybrid perovskites, our method suggests that the X-sites be populated by fluorine, rather than the usual iodine. A deeper investigation suggests that the reason for this is the enhanced stability of the fluorinated structures: while fluorinated structures have an average band gap of $\sim3$ eV and $\langle H_{\text{form}}\rangle \approx -1.1$ eV/atom, iodined perovskites have an average band gap of 2.6 eV, but a much higher $\langle H_{\text{form}}\rangle \approx -0.3$ eV/atom. 

\begin{figure}[!htb]
    \centering
    \includegraphics[width=1\textwidth]{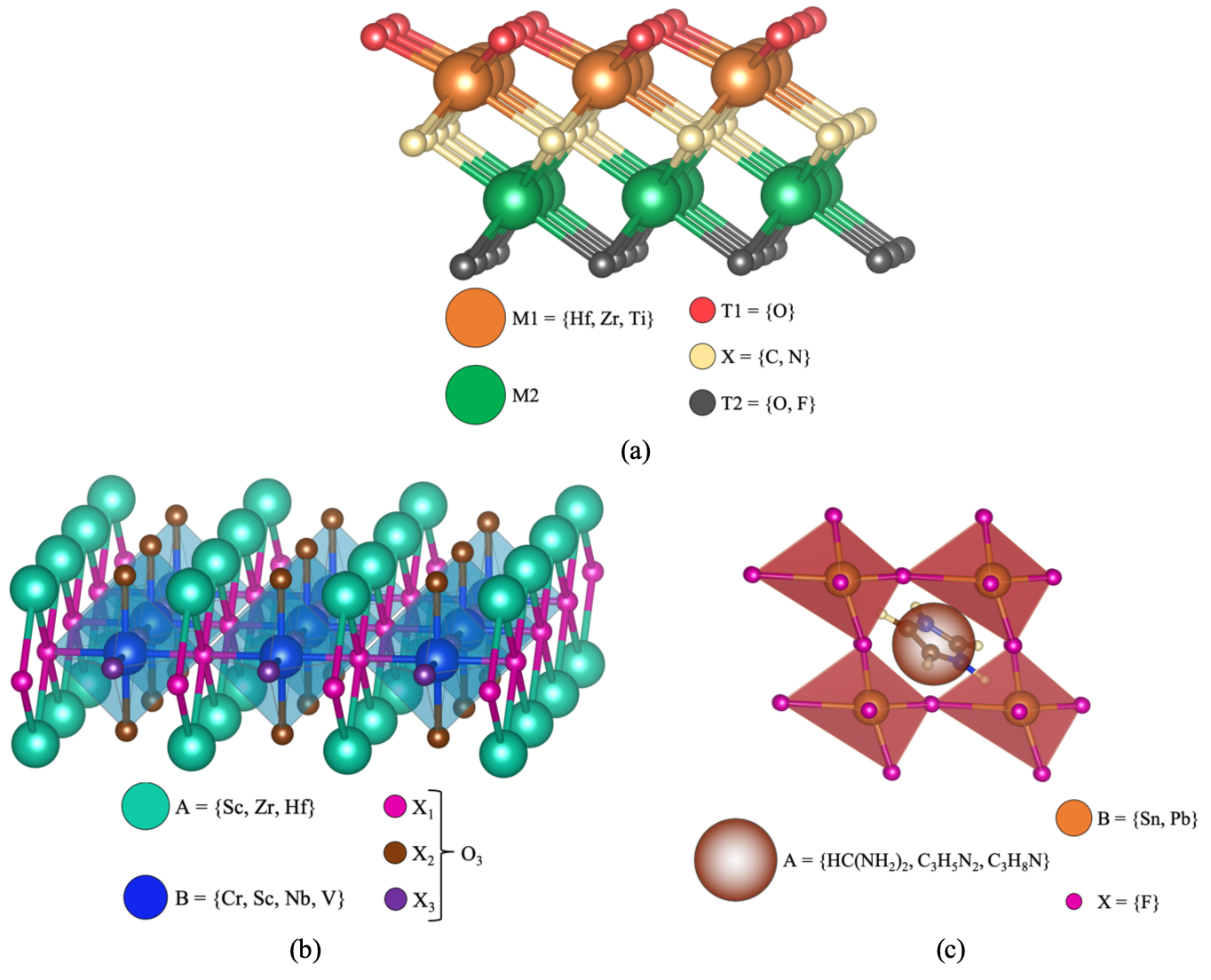}
    \caption{Summary of uncovered design rules for: (a) MXenes, where hafnium (Hf), zirconium (Zr), or titanium (Ti) must be bonded to an oxygen termination, regardless of the central X component or the metal M2 on the opposite side of the structure, which should have either an oxygen or a fluorine termination; (b) inorganic perovskites, in which the A-site must be occupied by scandium (Sc), zirconium (Zr), or hafnium (Hf), the B-site has to contain chromium (Cr), scandium (Sc), niobium (Nb), or vanadium (V), and, as discussed in the main text, all X-sites should be occupied by oxygen; (c) organic-inorganic hybrid perovskites, where all X-sites must be occupied by fluorine, the B-site can be occupied by either tin (Sn) or lead (Pb), and the A-site possibilities are fomamidinium \ce{(HC(NH2)2)}, imidazolium (\ce{C3H5N2}), and azetidinium(\ce{C3H8N}). Note: in the inorganic perovskite case, even though all X-site occupations should be the same, we represented them by different colors for completeness, since changing the value of $c_{\text{cut}}$ used in the analysis allows for other possibilities where all three distinct types of X-sites can be populated by a different atom.}
    \label{fig:dps_combined}
\end{figure}

Finally, for purely inorganic perovskites, we find that the A-sites should be occupied by scandium (Sc), hafnium (Hf), or zirconium (Zr). When analyzing the atomistic features used by the CGCNN model for these elements, we find that all of them have a covalent radius of $\sim 170$ pm, a first ionization potential in the neighborhood of 640 kJ/mol, and a 1.3 electronegativity in Pauling units. At the same time, the B-sites should be populated with vanadium (V), niobium (Nb), or chromium (Cr), all with atomic radii of $\sim 130$ pm, first ionization potential of roughly 650 kJ/mol, and an electronegativity of approximately 1.6 in the Pauling scale. Surprisingly, in the field of all-inorganic perovskites for photovoltaics applications, none of these compositions has been deeply investigated; focus has been more directed towards caesium-lead systems (\ce{CsPbX3}, where X can be I, Br, or Cl)\cite{ouedraogo2020stability}, indicating our work may contain new potential directions for further research in this area of science. Figure \ref{fig:dps_combined} contains a graphical summary of the top design principles uncovered in this work.

\begin{table}[htbp]
    \centering
    \begin{tabular}{c||ccccccc}
        \hline \hline
        & \makecell{Design\\ principle} & $c_{\text{DP}}$ & \makecell{$P_{\text{DP}}$\\\text{[\%]}}& \makecell{$P_{\text{DP|best}}$\\\text{[\%]}}& \makecell{$P_{\text{best|All}}$\\\text{[\%]}}& \makecell{$P_{\text{best|DP}}$\\\text{[\%]}}&
        \makecell{$P_{\text{DP|best}}$\\\hline $P_{\text{DP}}$}\\ 
        \hline\hline
        \multirow{5}{*}{\rotatebox[origin=c]{90}{MXenes}}
        & Hf*O-O-N & 0.729 & 0.046 & 12.069 & 0.243 & 63.636 & 261.897\\
        & Ti*O-O-N & 0.683 & 0.046 & 10.345 & 0.243 & 54.545 & 224.483\\
        & Hf*O-O & 0.560 & 0.092 & 18.966 & 0.243 & 50.000 & 205.776\\
        & Zr*O-F-C & 0.448 & 0.046 & 8.621 & 0.243 & 45.455 & 187.069\\
        & Hf*O-F & 0.570 & 0.092 & 17.241 & 0.243 & 45.455 & 187.069\\
        \hline\hline
        \multirow{5}{*}{\rotatebox[origin=c]{90}{\makecell{Inorganic \\Perovskites}}}
        & A=Sc-B=Cr & 0.264 & 0.037 & 10.526 & 0.100 & 28.571 & 284.632\\
        & A=Zr-B=Sc & 0.406 & 0.037 & 10.526 & 0.100 & 28.571 & 284.632\\
        & A=Sc-B=V & 0.320 & 0.037 & 5.263 & 0.100 & 14.286 & 142.316\\
        & A=Sc-B=Nb & 0.274 & 0.037 & 5.263 & 0.100 & 14.286 & 142.316\\
        & A=Hf-B=V & 0.294 & 0.037 & 5.263 & 0.100 & 14.286 & 142.316\\
        \hline\hline
        \multirow{5}{*}{\rotatebox[origin=c]{90}{\makecell{Organic\\Perovskites}}}
        & A=\ce{HC(NH2)2}-X=F & 0.817 & 0.446 & 13.333 & 2.229 & 66.667 & 29.911\\
        & A=\ce{C3H5N2}-X=F & 0.790 & 1.560 & 43.333 & 2.229 & 61.905 & 27.775\\
        & A=\ce{C3H8N}-X=F & 0.629 & 2.675 & 30.000 & 2.229 & 25.000 & 11.217\\
        & A=\ce{C3H5N2}-B=Sn & 0.282 & 2.452 & 26.667 & 2.229 & 24.242 & 10.877\\
        & A=\ce{C3H5N2}-B=Pb & 0.255 & 1.783 & 16.667 & 2.229 & 20.833 & 9.347\\
        \hline\hline
    \end{tabular}
    \caption{\label{tab:DP_table_res} Design principles with five highest values of ratio $\nicefrac{P_{\text{DP|best}}}{P_{\text{DP}}}$. The cutoff $c$-values used in computing values displayed for MXenes was $c_{\text{cut}}=0.95$, and for both inorganic and organic perovskite cases, $c_{\text{cut}}=0.80$.}
\end{table}

The collection of design rules we uncovered shows the capability of our model to both identify established criteria to attaining material performance, as well as to find new, unexplored avenues for application-focused material discovery, since it can be considered a basis for reverse engineering of 2D structures. We believe that machine learning methods such as CGCNN, coupled with a study of structural and compositional design rules, can open up paths for material innovation in a myriad of fields, including photovoltaics, electrochemistry, batteries, mechanical robust materials, among others. In the interest of further accelerating the discovery and screening of more 2D monolayer materials, we have open-sourced our code base on \href{https://github.com/victorventuri/cgcnn}{GitHub}.

 \section{Conclusions}
 Using fast and accurate graph convolutional neural networks on 2D materials, we screened large combinatorially generated datasets of MXene and perovskite materials in search of those with high likelihood of having properties of interest, as determined by the ensemble of trained CGCNN models. Using the results from the screening process, we were able to uncover the underlying design principles that make said structures useful candidates for their respective applications, which can be used guidance for both experimental and computational testing at different confidence levels.
 
 Furthermore, our design rules can act as a generative model for generating datasets by populating a series of structural sites with a given list of atoms. The metrics we adopted for evaluating design principles could prove most useful when dealing with combinatorically generated databases, typically the case for high-throughput screening. For instance, while $c_{\text{DP}}$ indicates how generally good a design rule is, the ratio $P_{\text{DP|best}}/P_{\text{DP}}$ shows how the knowledge that a candidate satisfies the specified design principle impacts its likelihood of being useful.

\section{Data availability}
The CGCNN modified code base can be found on \href{https://github.com/victorventuri/cgcnn}{GitHub}, as well as instructions on how to download it, set up a virtual environment, and run it. Further details that support the findings of this study are available from the corresponding author, V. Viswanathan, upon reasonable request.

\begin{acknowledgement}
The authors thank T. Xie \& R. Kurchin for helpful discussions. This work was supported in part by the Advanced Research Projects Agency-Energy Integration and Optimization of Novel Ion Conducting Solids (IONICS) program under Grant No. DEAR0000774. V. Venturi was supported in part by the Richard King Mellon Foundation Presidential Fellowship in Energy from the College of Engineering at Carnegie Mellon University. H. P. acknowledges support from the Office of Naval Research under Award No. N00014-19-1-2172. Acknowledgment is also made to the Extreme Science and Engineering Discovery Environment (XSEDE) for providing computational resources through Award No. TG-CTS180061.
\end{acknowledgement}

\begin{suppinfo}
The following files are available free of charge.
\begin{itemize}
\item suppinfo.pdf: Details on CGCNN hyperparameter optimization and ensemble prediction performance.

 \end{itemize}

 \end{suppinfo}

\bibliography{refs}


\section*{Supplementary material}

# Supporting Information: Machine Learning Enabled Discovery of Application Dependent Design Principles for Two-dimensional Materials

Victor Venturi,[†] Holden Parks,[†] Zeeshan Ahmad,[†] and Venkatasubramanian Viswanathan[*,†,‡]

[†]Department of Mechanical Engineering, Carnegie Mellon University, Pittsburgh, Pennsylvania 15213, USA

[‡]Department of Physics, Carnegie Mellon University, Pittsburgh, Pennsylvania 15213, USA

E-mail: venkvis@cmu.edu

# CGCNN Network Optimization

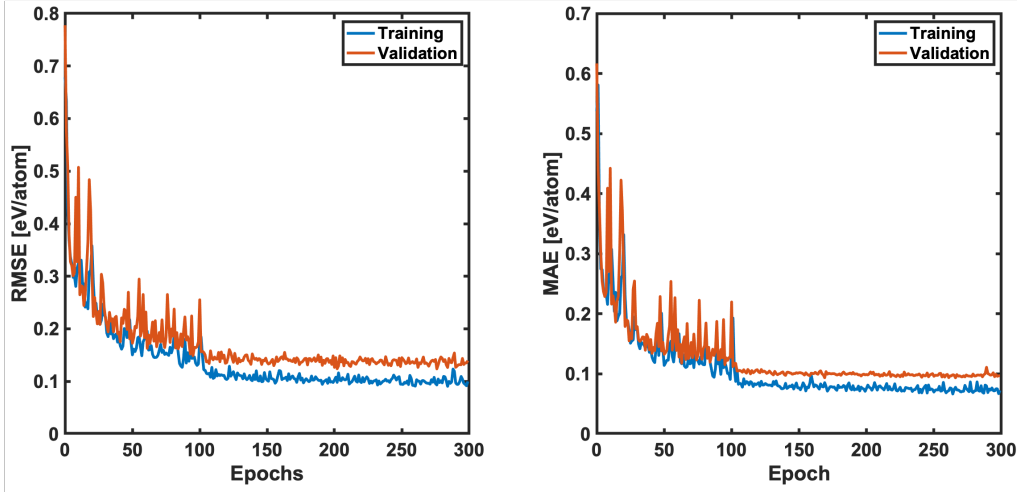

Figure S1: Example of root mean square (RMSE) and mean absolute error (MAE) curves used in optimization of network architecture and hyperparameters. These curves were used to evaluate the prediction performance of models using different number of convolution layers, hidden layers, epochs, number of neighbors used in convolution operations, among others. Represented here, we have the results from training a model to predict $H_{\text{form}}$ using a mean pooling function, 300 epochs, 1 hidden and 2 convolution layers.

In this work, we apply CGCNNs power of accurately predicting properties of periodic materials to investigate 2D materials, namely, MXenes and perovskites. Using a 70:15:15 training:validation:test split ratio on the C2DB database for the heat of formation property ($H_{\text{form}}$), we first optimized the network architecture, including number of convolution and hidden layers, learning rate, and number of epochs to be used in training, and the models' performances were evaluated as shown in Figure S1. We also tested different possibilities of pooling functions (mean, max, and min), as shown in Figure S2. The final architecture used in the models was composed of 2 convolutional layers and 1 hidden layer post-pooling. The networks were trained with a learning rate of 0.01 and a mean pooling function over 300 epochs.

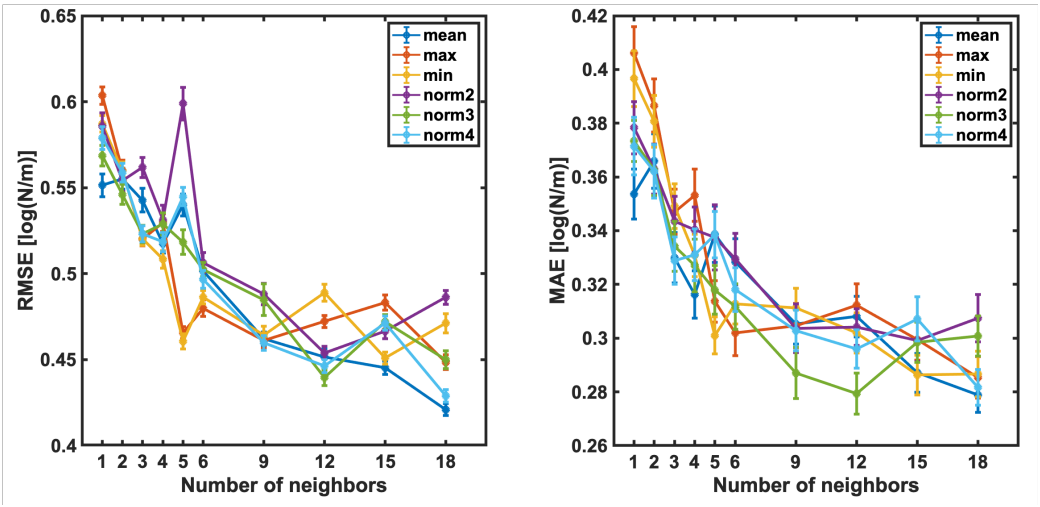

Figure S2: Evaluation of how pooling functions, as well as number of neighbors used in convolution operations, affect RMSE and MAE in $\log(c_{11})$ prediction. We also implemented higher order norm functions as pooling operators.

## CGCNN Ensemble Performance

In order to measure the performance of our model ensemble, we apply some of the metrics introduced by Kuleshov et al.[1] and,[2] namely, calibration and sharpness, besides mean absolute error (MAE) and root mean square error (RMSE).

In their work, the authors institute the concept of a calibration plot, which compares, for each predicted data point, the standard deviation of the ensemble predictions (y-axis) and the residual between the mean of the predictions and the true value of the data point – the mean error of the predictions (x-axis). In the regions where the observed estimation interval is greater than the expected interval (green), the model ensemble is called underconfident, since the true value falls within the error bars of the ensemble prediction. On the other hand, in the regions where the observed estimation interval is smaller than the expected interval (red), the model is called overconfident, since the standard deviation of the ensemble predictions do not encompass the true data value. The calibration plot for our 100 model ensemble trained to predict conduction band maximum is shown in Figure S3.

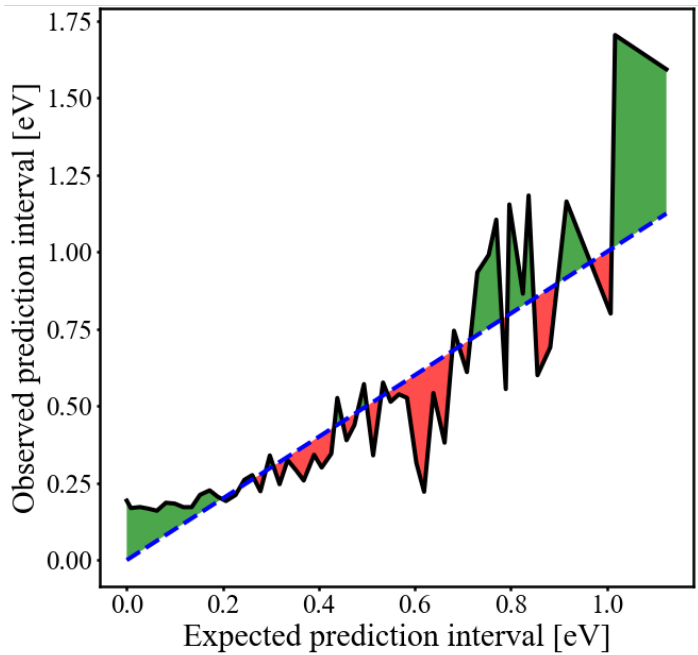

Figure S3: Calibration plot[1,2] of CGCNN ensemble CBM prediction. In general, the ensemble of model predictions captures the true values of CBM within one standard deviation.

However, measuring calibration is not sufficient, though necessary, for useful uncertainty quantification. For instance, a well calibrated model with large uncertainty estimates is less useful than a similarly calibrated model with smaller uncertainties. Thus, the concept of sharpness is introduced: a sharper model is that whose prediction standard deviations are smaller. This metric is measured in the following manner:[2]

$$\text{sharpness} = \sqrt{\frac{1}{N_{\text{dataset}}} \sum_{\text{structure}} Var[\mathcal{M}(\text{structure})]}$$

Figure S4 illustrates histograms of the mean prediction errors (red) and ensemble standard deviations (blue) of the models trained over band gap data. The values of prediction mean absolute error (MAE) and root mean square error (RMSE), as well as ensemble sharpness, are also represented. Table S1 contains some of these metrics for all predicted properties. Taking as example our band gap and heat of formation predictions, one can see

that our approach performs better than some first-principle simulations: for band gap, the accepted DFT errors are between 0.25 and 0.4 eV,[3,4] while both our MAE and sharpness fall on the lower end of this range; for $H_{\text{form}}$, DFT errors are of usually 0.1 eV/atom,[5] while all our uncertainty metrics are below this value by a safe margin.

CGCNN is a direction-agnostic framework for machine learning and since the material stiffness only depends on the relative positions of the atoms in the crystal, we can use them to predict the elastic constants as seen from the low MAE and RMSE (Table S1). We regress over the log of $c_{11}$ and $c_{22}$ since they are positive and want to avoid overweighing elastic constants of very stiff materials.

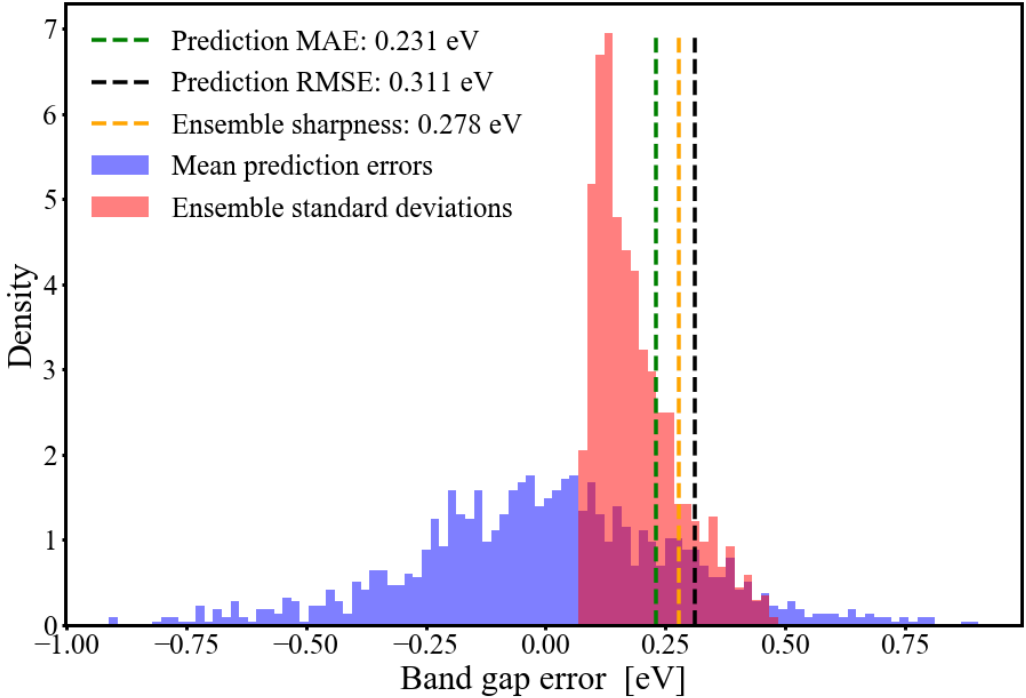

Figure S4: Histograms of errors on band gap prediction and of standard deviation of ensemble predictions.

Table S1: Metrics for uncertainty quantification of model ensembles

| Property | MAE | RMSE | Sharpness | Unit |
|---|---|---|---|---|
| $\log(c_{11})$ | 0.182 | 0.263 | 0.188 | $\log(N/m)$ |
| $c_{12}$ | 8.241 | 12.497 | 14.079 | N/m |
| $\log(c_{22})$ | 0.174 | 0.250 | 0.172 | $\log(N/m)$ |
| CBM | 0.193 | 0.264 | 0.310 | eV |
| VBM | 0.180 | 0.251 | 0.286 | eV |
| Band gap | 0.231 | 0.311 | 0.278 | eV |
| $H_{\text{form}}$ | 0.066 | 0.090 | 0.072 | eV/atom |
| Speed of sound x | 385.703 | 552.147 | 366.810 | m/s |
| Speed of sound y | 372.015 | 548.619 | 351.624 | m/s |



\end{document}